\documentclass[twocolumn,showpacs,showkeys,floatfix]{revtex4}

\usepackage{graphicx}
\usepackage{dcolumn}
\usepackage{amsmath}

\def\tvecr{{\mbox{\boldmath{$r$}}}}
\def\tvecp{{\mbox{\boldmath{$p$}}}}
\def\tvecx{{\mbox{\boldmath{$x$}}}}
\def\tvecy{{\mbox{\boldmath{$y$}}}}
\def\tvecxi{{\mbox{\boldmath{$\xi$}}}}
\newcommand{\xbld}[1]{\mbox{\boldmath $ #1 $}}

\def\rmd{{\rm d}}

\newcommand{\bra}{\langle}
\newcommand{\ket}{\rangle}

\begin{document}

\title{
Pauli-blocking Effect in a Quark Model
}
\author{Sachiko Takeuchi}
\affiliation{%
Japan College 
of Social Work, Kiyose, Japan}
\author{Kiyotaka Shimizu}
\affiliation{%
Department of Physics, Sophia University, 
Chiyoda-ku, Tokyo, Japan}

\date{\today}%

\begin{abstract}
Pauli-Blocking effect on the kinetic term is investigated
by employing the 
quark cluster model. 
The effect 
can be understood by 
the change of the degrees of the mixing between the
incoming wave and the 0$\ell$ state of the inter-cluster
wave function, which can be expressed by 
a potential which is highly nonlocal. 
We  look into
the properties of this effect by comparing 
equivalent local potentials.
In the channel where the Pauli-blocking effect is 
small, the on-shell equivalent local potential
simulates the nonlocal potential well even for the off-shell behavior. 
On the other hand, 
the off-shell behavior is 
very different from the original one where the effect is large.
This off-shell behavior, however, can well be simulated
by considering the nonlocal matrix elements only between 
the $0s$ state and the other states.
The energy dependent potentials
are also 
constructed and found to be helpful to understand the 
energy dependence of the effect.
\end{abstract}


\maketitle

\section{Introduction}

The cluster structure can often be seen in 
many-body systems.
One of the famous examples is 
$^{8}$Be, where the $^{4}$He  structure has
an important role \cite{cluster}.
It is also known,
that the quark cluster model, where the baryon
is considered as a cluster of three valence quarks,
reproduces successfully many of the features of 
the two-baryon systems \cite{oka,fae,oy,shi,osysup}.
In this work, we focus our attention 
on the Pauli-Blocking effect on the kinetic term.
We take the quark cluster model as an example
though the definition, the procedure, and the basic feature
discussed here can be 
applied to other cluster systems.

The Pauli-Blocking effect on the kinetic term 
should always be taken into account regardless of the 
interaction between the constituent fermions.
In this sense, this can be considered as a
`pure' Pauli-Blocking effect on a concerning system.
Actually,
as was reported in ref.\ \cite{TS02},
 in the channel where
the Pauli-blocking effect plays an important role,
the effect can still be seen in the phase shifts
after the 
quark interactions are introduced.

The low-energy properties of 
the quark Pauli-blocking effect
has been considered to cause
a strong repulsion between two baryons
in certain channels.
It, however, also produces an attractive force
for a higher energy region as we will discuss in detail
in this paper.
The Pauli-blocking effect among three or more clusters
may become important \cite{TS86}. Here, however, we concentrate
on the two-cluster systems.

In this paper, we investigate the Pauli-blocking effect 
by constructing a nonlocal potential
and by looking into 
the phase shifts and the wave functions of the systems
with this potential.
Then, we further investigate
its properties by comparing 
four kinds of local potentials:
1) the one which gives the same scattering phase shift as that of the 
original nonlocal potential,
2) the one which gives the same Born phase shift,
3) the energy-dependent potential obtained from the wave 
function,
and 4) the energy-dependent potential obtained by using the WKB 
method.

The contents of this article is as follows.
In section \ref{sec2}, we define the nonlocal potential
which expresses the Pauli-blocking effect on the nonrelativistic
kinetic term.
For this purpose, we employ the quark cluster model
without any interaction between quarks.
We will show that the effect can be taken into account 
by introducing a nonlocal two-baryon potential
which does not depend on the energy.

In section \ref{sec3}, 
we explain local potentials which express
a part of the features of the nonlocal potential.
In section \ref{sec31}, we mention that 
one can construct a unique potential which 
has the same on-shell behavior as that of a
given nonlocal potential.
In section \ref{sec32}, we construct a local potential 
which reproduces the same Born phase shift 
as that of the nonlocal potential.
For this local potential, the analytic form can be obtained.
In the next two subsections,
 two kinds of the energy-dependent local potentials
are discussed.
We mention that 
a potential can be obtained by a simple division 
by the wave function
in section \ref{sec33}.
The energy-dependent potential can also be 
derived with the help the WKB method, 
which is discussed in section \ref{sec34}.

The results are shown in section \ref{sec4}.
In section \ref{sec41}, 
we show that 
the  Pauli-Blocking effect can be expressed in terms of 
a nonlocal  potential, which is separable-like. 
It is found that the effect
does not cause a simple attraction or repulsion.
The phase shift changes its sign at the energy which corresponds to 
the kinetic energy of the relative $0\ell$ state.
As seen in section \ref{sec42},
the local potential which gives the same on-shell behavior
in the strongly prohibited channel
is found to have a deep attractive pocket at the very short range.
In the channel where the Pauli-blocking effect is 
small, or in the higher partial-wave channels,
this equivalent-local potential approach
seems valid even for the off-shell behavior;
 the wave function is also similar to that of the original nonlocal potential.
We find that the nonlocal treatment of the $0s$-$1s$ component
is essential to reproduce the off-shell behavior in the strongly
prohibited channel, which is discussed in section \ref{sec44}.

In section \ref{sec45},
the energy dependence of the potentials
is investigated.
The energy-dependent potentials obtained by the division by the wave function 
have singularities at the node points of the wave function.
They, however, are helpful to understand the 
energy dependence of the short range region.
The energy dependent potentials given
by the WKB method changes considerably as the energy increases.
The potential barrier in the strongly prohibited channel disappears
above a certain energy.
At the very high energy, all the potentials become gradually smaller.

Summary is given in section \ref{sec5}.

\section{Nonlocal potential arising from Pauli-blocking Effect}
\label{sec2}

The nonlocal potential arising from Pauli-blocking effect can be obtained as follows.
The Shr\"odinger equation for the constituent fermions (in the 
present problem, quarks) is
\begin{equation}
(H_{q} - E) \Psi = 0
\end{equation}
\begin{equation}
H_q = \sum_{i} \left( m_i + {\tvecp_i^2 \over 2 m_i} \right) -{\tvecp_G^2 \over 
2 m_G}+V_{q},
\end{equation}
where $m_i$ and $\tvecp_i$ are the mass and three momentum 
of the $i$th quark, respectively.
In the quark cluster model, 
the wave function is assumed to have the following form
 \cite{oka,fae,oy,shi,osysup}:
\begin{equation}
\Psi = {\cal A}_q \{\left[\phi_\alpha(b,\tvecxi_A) 
\phi_{\alpha'}(b,\tvecxi_B)\right] \chi(\tvecr_{AB})\}
\end{equation}
where ${\cal A}_q$ is the antisymmetrization operator on the six quarks,
$\phi_\alpha(b,\tvecxi)$ is the single baryon wave function with the quantum 
number $\alpha$, whose orbital part is the gaussian function 
 with the size parameter $b$.
By integrating the internal degrees of freedom out, 
we have the RGM (resonating group method) equation:
\begin{equation}
\int \left\{H(\tvecr,\tvecr')-E N(\tvecr,\tvecr')\right\}\chi(\tvecr') 
\rm{d}^3\tvecr'=0.
\label{eq:RGM}
\end{equation}
where $H$ and $N$ are the Hamiltonian and  normalization kernels:
\begin{widetext} 
\begin{eqnarray}
\left\{ \begin{array}{c}
H(\xbld{r},\xbld{r}')\\
N(\xbld{r},\xbld{r}')
\end{array} \right\}
= \int d\xbld{\xi}_Ad\xbld{\xi}_Bd\xbld{r}_{AB} 
\phi^\dagger_A(b,\xbld{\xi}_A)\phi^\dagger_B(b,\xbld{\xi}_B)
\delta(\xbld{r}-\xbld{r}_{AB})
\left\{ \begin{array}{c}
H\\ 1
\end{array} \right\} 
{\cal A} [\phi_A(b,\xbld{\xi}_A)\phi_B(b,\xbld{\xi}_B)
\delta(\xbld{r}'-\xbld{r}_{AB})]~.
\end{eqnarray}
\end{widetext}
This equation can be rewritten in the Sch\"odinger form as 
\begin{equation}
\int \left\{{\overline H}(\tvecr,\tvecr')-E\right\}{\overline \chi(\tvecr')} \rm{d}^3\tvecr'=0.
\end{equation}
The new kernel ${\overline H}$ and the wave function ${\overline \chi}$ are defined as
\begin{eqnarray}
{\overline H}
&=&N^{-1/2}HN^{-1/2} \label{eq:Hbar}\\
{\overline \chi}
&=&N^{1/2}\chi . \label{eq:chibar}
\end{eqnarray}
The potential for baryons is defined from the above kernel as:
\begin{equation}
V_{QCM} = {\overline H}-K_{D} .
\end{equation}
where $K_D$ is the usual kinetic term 
of the two baryons with the mass $\sum m_i$.
The Schr\"odinger equation with this potential,
\begin{equation}
(K_{D}+V_{QCM}-E)\psi = 0,
\label{eq:sch}
\end{equation}
can be treated as the one for the baryon system with implicit internal 
degrees of freedom\cite{TS02}.
There are other definitions to extract a two-baryon potential
from the RGM equation.
For example, $V \equiv H-EN-(K_D-E)$ can be also taken as 
a potential between baryons. It, however, depends on the energy 
rather strongly while the nonlocality is still as large as the present one
\cite{Fujiwara, Ta02}.
The present definition is a unique way to remove the energy dependence
from the two-baryon potential.

Suppose all of the particles cannot occupy the same 
orbital state at the same time because of the Pauli principle, the 
relative 0$s$ two-cluster state is forbidden.
When the system has such forbidden state(s), eqs.\ (\ref{eq:Hbar}) and 
(\ref{eq:chibar}) are
modified as
\begin{eqnarray}
{\overline H}
&=&(PNP)^{-1/2}(PHP)(PNP)^{-1/2} \label{eq:HbarP}\\
{\overline \chi}
&=&(PNP)^{1/2}P\chi  \label{eq:chibarP}
\end{eqnarray}
where, $P$ is the projection operator on the space of all allowed states.
The solution of the Schr\"odinger equation eq.\ (\ref{eq:sch})
should be orthogonal to the forbidden states.
Since eigenvalues of both of the $H$ and $N$ become zero for the forbidden states,
eqs.\ (\ref{eq:HbarP}) and 
(\ref{eq:chibarP}) can be defined as a natural extension of eqs.\ 
(\ref{eq:Hbar}) and (\ref{eq:chibar})
as we will see later.

When we take only the nonrelativistic kinetic term as $H_{q}$, 
with no interaction 
between quarks, the hamiltonian kernel in eq.\ (\ref{eq:RGM})
can be replaced by the kinetic kernel, $K(\tvecr,\tvecr')$.
Then, the QCM potential becomes,
\begin{equation}
V_K = N^{-1/2}KN^{-1/2} - K_D~,
\label{eq:VK}
\end{equation}
which is considered to express 
the effect of the Pauli-blocking on the kinetic term.

In this paper, 
we focus our attention on the single channel which has no 
forbidden state but the normalization kernel deviated from 1.
Also, we take the two-cluster system where each cluster has
three fermions with a common value for the mass, $m_{q}$. 
In this case, we only have one exchange
term, provided that the relative wave function 
as well as the internal wave function of the clusters
are antisymmetrized.

The normalization kernel $N(\tvecr,\tvecr')$ and 
the kinetic kernel $K(\tvecr,\tvecr')$ in eq.\ (\ref{eq:RGM})
become
\begin{eqnarray}
N(\tvecr,\tvecr') &=& \delta(\tvecr-\tvecr')+N_{\rm ex}(\tvecr,\tvecr')
\end{eqnarray}
and
\begin{widetext} 
\begin{eqnarray}
K(\tvecr,\tvecr') &=& K_{D}+
K_{0}\left({15\over 2}-{27\over 16 b^{2}}(r^{2}+r'{}^{2})
+{21\over 8 b^{2}}\tvecr\cdot\tvecr'\right)
N_{\rm ex}(\tvecr,\tvecr')
\end{eqnarray}
respectively, where
$K_D$ is now the kinetic energy
of the two-baryon system with the reduced mass $3m_q/2$,
and $K_0=3/(4m_q b^2)$.
The exchange part of the normalization kernel is obtained as
\begin{eqnarray}
N_{\rm ex}(\tvecr,\tvecr') & = &(\nu-1)\;\left({27 \over 16\pi b^2}\right)^{3\over2}
\exp\left[-{15\over 16 b^{2}}(r^{2}+r'{}^{2})
+{9\over 8 b^{2}}\tvecr\cdot\tvecr'\right]~.
\label{eq:nu}
\end{eqnarray}
(See appendix.)
The factor $(\nu-1)$ is the matrix element of the exchange operator 
in the color-spin-flavor space,
$-\bra \sum P_{ij}\ket$.
In our treatment here the model also
includes the spectroscopic factor, which comes essentially from the difference in the 
number of combinations for picking up a cluster 
(three quarks out of six instead of one baryon out of two).
Thus, the value of $\nu$ can be more than 1;
actually, it is 0 or 10 when the states are
totally antisymmetrized.
In the following calculation,
we take $\nu=10/9$ or 2/9 as an example,
which are typical values for the single channel two-baryon systems.

Both of $N$ and $K$
can be expanded by the harmonic oscillator wave function 
with the size parameter $\beta=\sqrt{{2\over 3}}b$ as:
\begin{eqnarray}
N(\tvecr,\tvecr') &=& \sum_{n\ell m} \left(1+ (\nu-1)\,\theta^{2 n + \ell} \right) 
\psi_{n\ell m}(\beta,\tvecr)\psi_{n\ell m}^*(\beta,\tvecr')
\end{eqnarray}
and
\begin{eqnarray}
K(\tvecr,\tvecr') &=& \sum_{nn'\ell m} {2K_{0}\over 3}\left\{
\delta_{nn'}
\left(2 n + \ell + {3\over2} \right)\left(1+(\nu-1)\,\theta^{2 n + \ell} 
\right)\right. \nonumber\\
&+&
\left.(\delta_{nn'+1}+\delta_{nn'-1})
\sqrt{(n_{<}+1)\left(n_{<} + \ell + {3\over2} \right)}
\left(1+(\nu-1)\,\theta^{2 n_{<} + \ell} \right) 
 \right\}
\psi_{n\ell m}(\beta,\tvecr)\psi_{n'\ell m}^*(\beta,\tvecr') 
\end{eqnarray}
where 
$\theta=1/3$,  
$\psi_{n\ell m}$ is the harmonic oscillator wave function of the 
quantum number $n \ell m$, and
$n_{<}$ corresponds to the smaller one among $n$ and $n'$.

Thus the obtained $V_K$ can be expressed as
\begin{eqnarray}
V_K(\tvecr,\tvecr') &=& \sum_{nn'\ell m} V_{K}^{nn'\ell}
\psi_{n\ell m}(\beta,\tvecr)\psi_{n'\ell m}^*(\beta,\tvecr') 
\label{eq:VKrr}
\end{eqnarray}
with
\begin{eqnarray}
V_{K}^{nn'\ell} &=& 
(\delta_{nn'+1}+\delta_{nn'-1}) {2K_{0}\over 3}
\sqrt{(n_{<}+1)\left(n_{<} + \ell + {3\over2} \right)}\left
\{\sqrt{1+ (\nu-1)\,\theta^{2 n_{<} + \ell} \over 1+ {(\nu-1)\,\theta^{2 n_{<} + \ell+2}}}
-1
 \right\}
\label{eq:KV}
\end{eqnarray}
\end{widetext} 
As was mentioned before,
eq.\ (\ref{eq:KV}) is also valid even when $\nu=0$, namely, a forbidden
state exists, or when the system contains the coupled channels.

Note that the diagonal part, namely, the $n=n'$ term,
does not exist in $V_K$.
The diagonal parts in the original $K(\tvecr,\tvecr')$ kernel
are canceled out when $K$ is divided by $N$ and subtracted by 
$K_D$. 

For the partial wave decomposition of the nonlocal potential below,
we use the notation:
\begin{equation}
V_K(\tvecr,\tvecr') = \sum_{\ell m} V_{K\ell}(r,r') Y_{\ell m}(\tvecr)Y^*_{\ell m}(\tvecr')~.
\end{equation}

Since the terms in eq.\ (\ref{eq:VKrr}) are of order $O((1/9)^n)$, 
only the small $n$ terms are important.
The radial part of the lowest-order term of eq.\ (\ref{eq:KV}),
where $n$ or $n'$ is zero, is written as
\begin{eqnarray}
V^{(0)}_{K\ell}(r,r') &=&
\left(\sqrt{\nu}
-1 \right) {2K_{\beta}\over 3} 
\sqrt{\ell+{3\over2}}
\nonumber\\
&\times&\left(
u_{0\ell}(\beta,r)u_{1\ell}(\beta,r')
+
u_{1\ell}(\beta,r)u_{0\ell}(\beta,r')
\right)~,
\label{eq:KV0}
\end{eqnarray}
where $u_{n\ell}$ is the orbital part of the harmonic oscillator wave 
function, $\psi_{n\ell m}$.
>From the above equation, it is clearly seen 
that the sign and magnitude
of the potential changes according to those of $(\nu-1)$.

Since $r$- and $r'$-dependence of the factor in eq.\ (\ref{eq:KV0}) 
is 
\begin{eqnarray}
\lefteqn{u_{0\ell}(\beta,r)u_{1\ell}(\beta,r')
+
u_{1\ell}(\beta,r)u_{0\ell}(\beta,r') ~\propto} \nonumber\\
&&
{2\over \beta^{3}}
\left({r r'\over \beta^{2}}\right)^{\ell}
\left(1-{1\over 2\ell+3}{(r^{2}+r'{}^{2}) \over \beta^{2}}\right)
\nonumber\\
&\times&\exp[-{(r^{2}+r'{}^{2})\over 2 \beta^{2}}]~,
\end{eqnarray}
the potential given by the lowest term 
has a rather simple structure.

For each of the angular momentum $\ell$, the phase shift given by $V_{K\ell}$ 
with the Born approximation can be written as:
\begin{eqnarray}
\lefteqn{\tan \delta^{Born}_{\ell}(k)} \nonumber\\ 
 &=&
-2 \mu k \int j_{\ell}(kr) V_{K\ell}(r,r') 
j_{\ell}(kr') \, r^{2}\rmd r \,  r'{}^{2}\rmd r' \nonumber\\ 
& = &
-2 \mu k \sum_{nn'} 
V_{K}^{nn'\ell}
\tilde{u}_{n\ell}(\beta,k)\tilde{u}_{n'\ell}(\beta,k)
\label{eq:delB}
\end{eqnarray}
with $k^{2}=2 \mu E$. The function 
$\tilde{u}_{n\ell}(\beta,k)$ 
is the Fourier transformation of the $u_{n\ell}(\beta,r)$, 
namely,
$\tilde{u}_{n\ell}(\beta,k) = \sqrt{\pi/2}(-1)^n u_{n\ell}(1/\beta,k)$.

The momentum which gives ${\delta(k_{0})=0}$,  $k_{0}$,
can be obtained as the solution of
$\tilde{u}_{n=1\ell}(\beta,k_{0})=0$ approximately.
This momentum $k_0$ 
depends on the cluster size parameter, $b$, 
but does not depend on $\nu$ nor other parameters.
When $0< \nu \ll 1$, the resonance becomes sharper,
but the resonance momentum does not change.
The resonance energy of the almost-forbidden channel
is given by $\beta^2 k_{0}^2 \sim (\ell+{3\over 2})$, 
which corresponds to the
 kinetic energy of the $0\ell$ state.

\section{Local potentials}
\label{sec3}
\subsection{On-shell-equivalent local potential}
\label{sec31}

A local potential can be 
constructed uniquely from a given $S$-matrix, $S(k)$,
 by using the 
Marchenko-method\cite{TS02,Newton,AM63,CS89}.
The procedure is as follows.
First,  the following function $F(r)$ should be calculated 
from the S-matrix with poles at \{$k=i\kappa_j$\},
\begin{equation}
F(r)=-\frac{1}{2\pi} \int_{-\infty}^{+\infty} \! e^{ikr} \{ S(k)-1 \} dk
+\sum_j c_j^2e^{-\kappa_j r},
\label{eq22}
\end{equation}
where $c_j^2$ is
\begin{equation}
c_j^2=\mbox{Residue} \{ S(k) \}
\mbox{ at } k=i\kappa_j~(\kappa_i>0).
\end{equation}
Next, we solve the following integral equation with $F(r)$.
\begin{equation}
K(r,r') = -F(r+r')-\int_r^{\infty} \! F(r+r'')K(r,r'') d r''. 
\end{equation}
Then, the local potential $V(r)$ is obtained from 
the solution of the above equation, $K$, as
\begin{equation}
2\mu V(r) = -2 \frac{d}{dr}K(r,r),
\label{eq25}
\end{equation}
where $\mu$ is the reduced mass of the system.
The more detailed calculation is found, for example, in 
ref.\ \cite{TS02}.

We construct a local potential from the 
phase shift given by the nonlocal potential $V_{K}$.
The obtained potential is the momentum- and energy-independent, 
but depends on the angular momentum $\ell$.
We call this the on-shell-equivalent local potential from now on.
Its on-shell behavior, namely the asymptotic behavior of the
 wave function obtained by solving the Schr\"odinger equation
 with that potential, is the same as 
that of the original nonlocal potential.
The off-shell feature, however, can be very different from 
the original one when the degree of the nonlocality is large.
The difference can be seen, for example, by looking into the
 wave functions at finite $r$.

For the system with a forbidden state,
one can also construct a potential which reproduces the phase shift.
When the wave function of the forbidden state 
behaves asymptotically for large $r$ as $\exp[-ar]$,
the on-shell-equivalent local potential is also uniquely
constructed from its binding energy and residue 
together with the phase shifts.
The wave functions obtained by solving this
local potential, however, are not orthogonal to the original
wave function of the forbidden state in general.
Moreover, in the present problem, the forbidden state
is bounded by the confinement force.
A local potential with a finite size cannot produce
the asymptotic behavior of its wave function, $\exp[-ar^2]$.

As we will show later, 
there are channels where the Pauli-blocking effect is strong
even though it does not produce a forbidden state.
There, the off-shell behavior of the on-shell-equivalent local potential 
is very different from the original one.
For that case, 
we consider the partially local potential,
where the $0s$-$1s$ component
is treated as nonlocal,
\begin{equation}
V^{0s{\rm -nonloc}}(r,r') = 
V^{(0)}_{K\ell}(r,r')
+V^{0s{\rm -nonloc}}_{loc}(r) {\delta(r-r')\over r^2}~.\label{eq:0snloc}
\end{equation}
There may be no existence nor uniqueness 
for the local part, $V^{0s{\rm -nonloc}}_{loc}$.
In the case we will describe later, however, 
we can find $V^{0s{\rm -nonloc}}_{loc}$ by fitting the phase shifts.

\subsection{Born-Equivalent local potential}
\label{sec32}

It 
is also useful to look into the local potential which can 
be expressed in an analytic form.
For a system with no bound state nor the forbidden state, 
we construct the local potential which
reproduces the same Born phase shift as that of the nonlocal 
potential. We call this  Born-equivalent local 
potential.
 
The Born-equivalent local potential is the same as 
the on-shell-equivalent local potential 
when the original potential is local, even where
 the Born approximation is not valid.
It, however, deviates from the on-shell-equivalent potential in general
when the original potential is nonlocal.
Thus, we consider the size of the deviation as 
a rough estimate of the size of the nonlocality.

Suppose the system does not have 
a bound state.
Then, the phase shift by the first Born approximation, 
which is the Fourier transformation 
of the potential, contains enough information to construct a
local potential.
For example, 
it is obtained uniquely  for $\ell=0$ by the cosine Fourier
transformation.
>From the known $\delta^{Born}_{\ell =0}(k)$,
\begin{eqnarray}
\lefteqn{\tan \delta^{Born}_{\ell =0}(k)}\nonumber\\ &=& 
-2 \mu k \int j_{0}(kr) V_{\ell =0}(r) j_{0}(kr)\,  r^{2}\rmd r 
\\
&=& { \mu \over k} 
\left(\int_{0}^{\infty} \! \cos(2kr)\,  V_{\ell =0}(r)\,  \rmd r
+ const.\right)~,
\end{eqnarray}
$V_{\ell=0}(r)$ is obtained as
\begin{eqnarray}
V_{\ell=0}(r) &=& {4\over \pi}\int_{0}^{\infty}
{k\over \mu} \tan \delta^{Born}_{\ell=0}(k)\, \cos(2kr)\, \rmd k
\end{eqnarray}
with an ambiguity of the Dirac's delta function at $r=0$, 
which does not contribute to the observables.

The explicit form for the $\ell=0$ system of the present issue, 
for example, can be obtained as
\begin{eqnarray}
\lefteqn{
V_{\ell=0}(r) 
=
 -{8\over \pi}
\sum_{nn' \atop \ell=m=0} 
V_{K}^{nn'\ell}
}&&
 \nonumber\\
&\times&
\int_{0}^{\infty} \tilde{u}_{n\ell}(k) \tilde{u}_{n'\ell}(k)
\, \cos(2kr)\,  k^{2}\rmd k
~.
\end{eqnarray}
The one which corresponds to $V^{(0)}_{K\ell=0}$ 
in eq.\ (\ref{eq:KV0}) is:
\begin{eqnarray}
\lefteqn{V^{(0)}_{K0~Born}  }\nonumber\\
 &=& 16 K_{0}\left(\sqrt{\nu}-1 \right)
{r^{2}\over \beta^{2}}\left(1-{2\over 3}{r^{2}\over\beta^{2}}\right)
\exp[-{r^{2}\over \beta^{2}}]~.\nonumber\\
\end{eqnarray}
The sign of this potential changes at $r\sim \sqrt{3\over 2}\beta =b$.

\subsection{Energy-dependent local potential (by direct division)}
\label{sec33}

An energy-dependent potential
can be obtained directly by dividing
the hamiltonian by the obtained wave function.
When $\psi(r)$ is a solution of the Schr\"odinger equation with 
a local potential $V(r)$, $V(r)$ can be reconstructed from the 
wave function:
\begin{equation}
V(r) = E-{1\over 2 \mu}{p^{2}\psi(r)\over \psi(r)}
\label{eq:vdiv}
\end{equation}

Suppose $\psi$ is the solution of 
the  Schr\"odinger equation with 
a nonlocal potential, $V(r)$ obtained from the above equation
can also be considered as an equivalent local potential, 
which depends on all of $\ell$, $r$, and $E$.

The potential obtained this way 
gives the same phase shift and the wave function 
as the original nonlocal potential.
The potential, however, 
may have a singularity of the order $1/(r-r_{0})$ 
at the node points of the 
wave function in general.
Nevertheless, we discuss this potential  
to show the energy dependence of the Pauli-blocking effect.
It has an advantage that one can always construct 
this potential except the node point of the wave function
unlike the one defined using WKB in the next subsection.

\begin{figure*}
	\includegraphics*[scale=0.6]{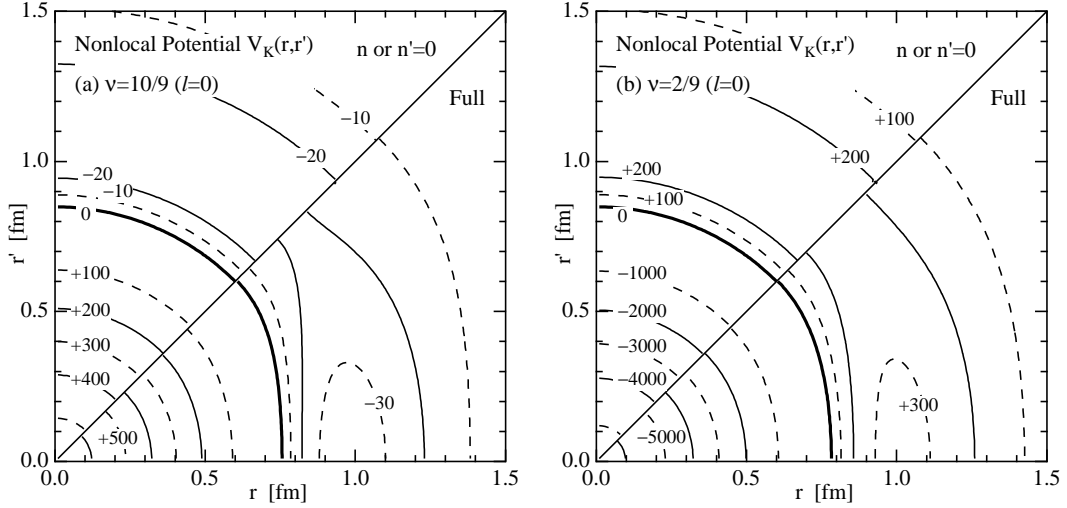}
	\caption{Contour plot of the
original nonlocal potential $V_{K \ell}(r,r')$
for the $\ell=0$ channel.
(a) is for the $\nu=10/9$ case,
and (b) is for the $\nu=2/9$ case.
In the upper half of each figure 
we plot the lowest-order term ($n {\rm ~or~} n'=0$),
defined in eq.\ (\ref{eq:KV0}).
	\label{fig01}}
\end{figure*}

\subsection{Energy-dependent local potential (by WKB method)}
\label{sec34}

One of the conventional ways to treat the nonlocality of the potential 
is to interpret it as the energy dependence by using the WKB method
\cite{suz}.

In this subsection, 
we explain the effect of the nonlocality and a relation
between the nonlocality and energy dependence.
The nonlocal part has the following form in the Sch\"{o}dinger equation.
\begin{equation}
\int \bra \tvecr \mid V \mid \tvecr' \ket
\chi_{\ell} (\tvecr') d \tvecr'
\end{equation}
Introducing $\xbld{s}=\tvecr'-\tvecr$, we obtain
\begin{eqnarray}
\lefteqn{\int \bra \tvecr \mid V \mid \tvecr' \ket
\chi_{\ell} (\tvecr') d \tvecr'}\nonumber\\
 &=& \int \exp (\frac{\xbld{\nabla} \cdot
\xbld{s}}{2})U(\tvecr,\xbld{s})\exp (\frac{\xbld{\nabla} \cdot
\xbld{s}}{2}) d \xbld{s} \; \chi_{\ell} (\tvecr)\nonumber\\
\label{eq:Edep1}
\end{eqnarray}
Here we have used the following relations:
\begin{eqnarray}
\chi_{\ell} (\tvecr')&=&\chi_{\ell} (\tvecr+\xbld{s}) =
\exp(\xbld{\nabla} \cdot \xbld{s}) \chi_{\ell} (\tvecr) 
\end{eqnarray}
and
\begin{eqnarray}
\lefteqn{\bra \tvecr \mid V \mid \tvecr' \ket =  U(\frac{\tvecr+\tvecr'}{2},
\xbld{s})
=U(\tvecr+\frac{\xbld{s}}{2},\xbld{s}) } &&\nonumber\\
&=&
\exp (\frac{\xbld{\nabla} \cdot \xbld{s}}{2})U(\tvecr,\xbld{s})\exp
(-\frac{\xbld{\nabla} \cdot \xbld{s}}{2}) \rule{1cm}{0cm}
\label{eq:EdepRL}
\end{eqnarray}

Assuming that $U(\tvecr,\xbld{s})$ does not change
rapidly within a distance $s$, we get a momentum dependent local potential
\begin{eqnarray}
\lefteqn{\int \bra \tvecr \mid V \mid \tvecr' \ket
\chi_{\ell} (\tvecr') d \tvecr' }\nonumber \\ &=& 
\int U(\tvecr,\xbld{s})\exp (\xbld{\nabla}
\cdot \xbld{s}) d \xbld{s}  \; \chi_{\ell} (\tvecr)
=U(\tvecr,\xbld{p}) \chi_{\ell} (\tvecr)~.\nonumber \\ 
\label{eq:Edep2}
\end{eqnarray}
Here $\xbld{\nabla}$ acts only on the wave function and thus replaced by
the momentum operator $\xbld{p}$.
The present potential is central and depends on $\xbld{p}$ and $\xbld{r}$ 
only through $p^2$, $r^2$ 
and $(\xbld{p}\cdot\xbld{r})^{2}$.
By 
substituting 
$(\xbld{p}\cdot\xbld{r})^{2}\rightarrow 
p^{2}r^{2} - (\ell+1/2)^{2}$, we have
the potential, $U(r,p)$, which depends only on $p^{2}$ and $r^2$.

Employing the classical approximation,
\begin{equation}
\frac{{p}^2}{2\mu}+U(r,p)=E \left(={k^2\over 2 \mu}\right)
\label{eq:p2E}
\end{equation}
we obtain the energy-dependent local potential,
$U(r,E)$ by solving eq.\ (\ref{eq:p2E})
for each $r$ and $E$
with the explicit form of $U(r,E)$.

For the $n$ or $n'=0$ term of the $\ell=0$ channel, we find that
 $U(r,p)$ has the form,
\begin{eqnarray}
U^{(0)}_{\ell=0}(r,p) &=& (\sqrt{\nu}-1) {32\over3} K_{\beta} 
(-{r^{2}\over \beta^{2}}+\beta^{2}p^{2})
\nonumber \\&\times&\exp[-\beta^{2}p^{2}]
\exp[-{r^{2}\over \beta^{2}}]~.
\end{eqnarray}

When $\beta^2 p^{2}$ is small, the potential becomes
\begin{eqnarray}
\lefteqn{U^{(0)}_{\ell=0}(r,E) \sim}  \nonumber\\
&&
{8  (\sqrt{\nu}-1) \over \mu \beta^{2}}
{
\left\{ -{r^{2}\over \beta^{2}} 
+ 2 E \mu \beta^{2} \left(1+{r^{2}\over \beta^{2}} \right)
\right\}
\exp[-{r^{2}\over \beta^{2}}]
\over
1+16 (\sqrt{\nu}-1)
\left(1+{r^{2}\over \beta^{2}} \right)
\exp[-{r^{2}\over \beta^{2}}]
}
~.\nonumber\\
\label{eq:U0rE}
\end{eqnarray}
For the $\nu>1$ case,
the potential is negative definite at $E=0$
and increases as the energy increases.
When $U$ is small compared to $E$,
then $U(r,E)$ decreases by the factor
$\exp[-2\mu\beta^2 E]$.

\section{Results}
\label{sec4}

\subsection{Nonlocal potential}
\label{sec41}

\begin{figure*}
	\includegraphics*[scale=0.6]{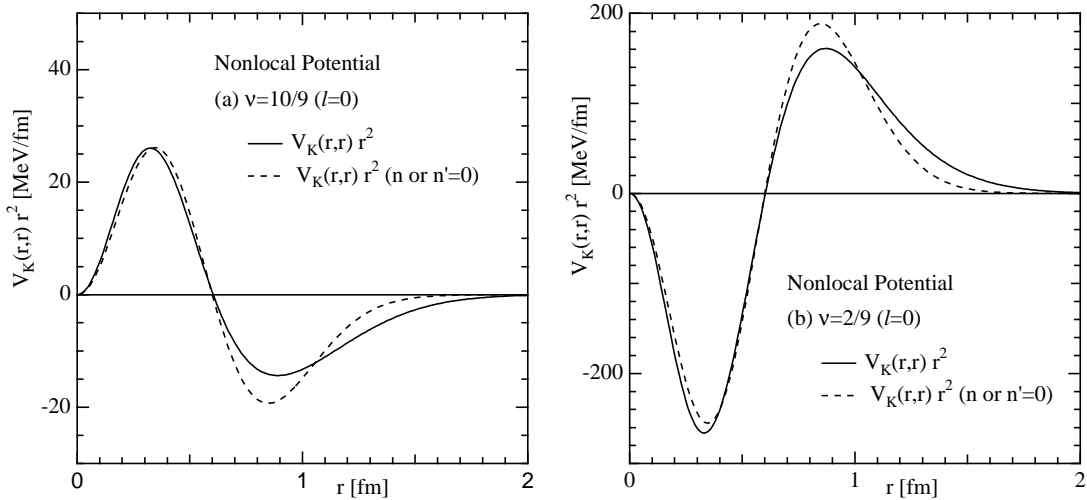}
	\caption{Diagonal part of the
original nonlocal potential,
$V_{K \ell}(r,r'\! =\! r)r^2$,
for the $\ell=0$ channel.
The solid lines are for the full calculation,
and the dotted lines are for the lowest-order term.
(a) is for the $\nu=10/9$ case,
and (b) is for the $\nu=2/9$ case.
\label{fig02}}
\end{figure*}

The nonlocal
potential, $V_{K}(r,r')$, defined by eq.\ (\ref{eq:VK})
and its diagonal part, $V_{K}(r,r'\! =\! r)r^{2}$,
are shown in Figure \ref{fig01} (contour plot) and in Figure \ref{fig02}, respectively,
for the $\ell=0$ channel.
The cluster size parameter here is $b =$ 0.6 fm, and 
the quark mass is taken to be $m_{q}=$ 313 MeV.
%
\begin{figure*}
	\includegraphics*[scale=0.6]{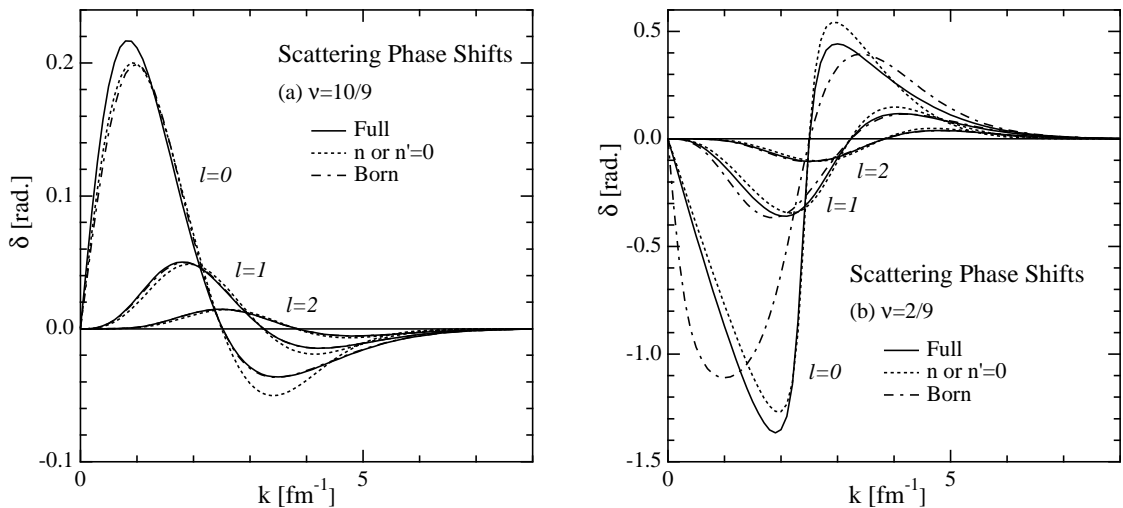}
	\caption{Scattering phase shifts obtained 
from the original nonlocal potential $V_{K \ell}$
for the $\ell=0$, 1 and 2.
The solid lines are for the full calculation,
the dotted lines are for the lowest-order term,
and the dot-dashed lines are the results from
the Born approximation.
(a) is for the $\nu=10/9$ case,
and (b) is for the $\nu=2/9$ case.
\label{fig1}
}
\end{figure*}
In Figures \ref{fig01}(a) and  \ref{fig02}(a), 
we plot the potential for the $\nu=10/9$ case, 
which corresponds to the even partial wave of the NN channel.
Overall size is small because $\nu$ is close to one.
Figures \ref{fig01}(b) and \ref{fig02}(b) correspond to the 
$\Sigma$N($I$=3/2,$S$=1) channel, where the  Pauli-blocking effect is 
strong ($\nu=2/9$).
There the size is very large; it is about 7/9 times as large as 
the kinetic energy of the 
$0\ell$ state in the short range region.

In the contour plots, we plot the potential given by the lowest term,
$V^{(0)}_{K\ell}(r,r')$ in eq.\ (\ref{eq:KV0}),
in the upper halves.
The potential $V^{(0)}_{K\ell=0}$
depends only on $(r^{2}+r'{}^{2})$ for the $\ell=0$ channel.
 The full potentials, written
in the lower halves, however, 
have more complicated structure around $r\sim $ 1 fm.

\begin{figure*}
	\includegraphics*[scale=0.6]{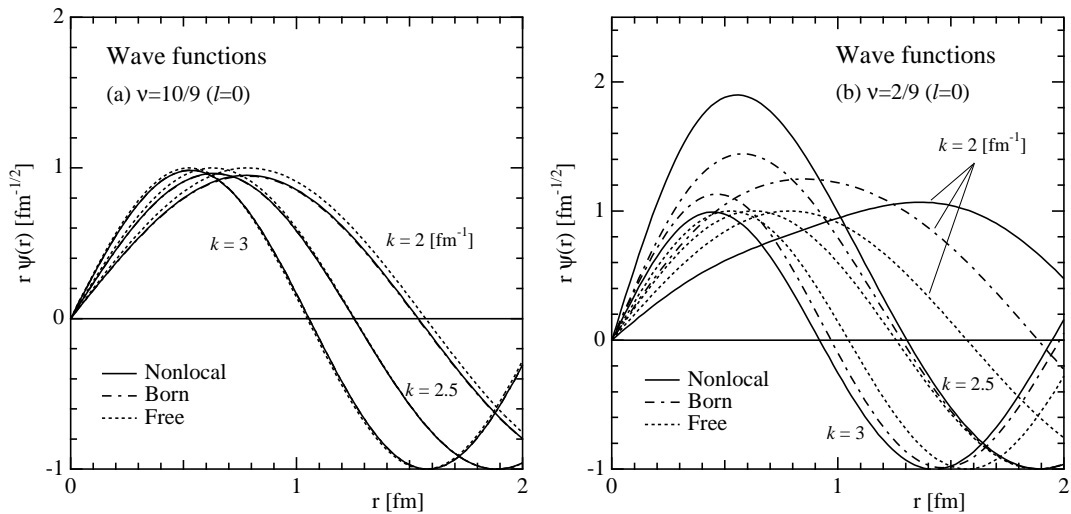}
	\caption{Wave functions obtained 
from the original nonlocal potential $V_{K \ell}$
for the $\ell=0$ channel.
The dotted lines are the free wave function,
and the dot-dashed lines are the results from the 
Born approximation.
(a) is for the $\nu=10/9$ case,
and (b) is for the $\nu=2/9$ case.
\label{wfn1}
}
\end{figure*}
\begin{figure*}
	\includegraphics*[scale=0.6]{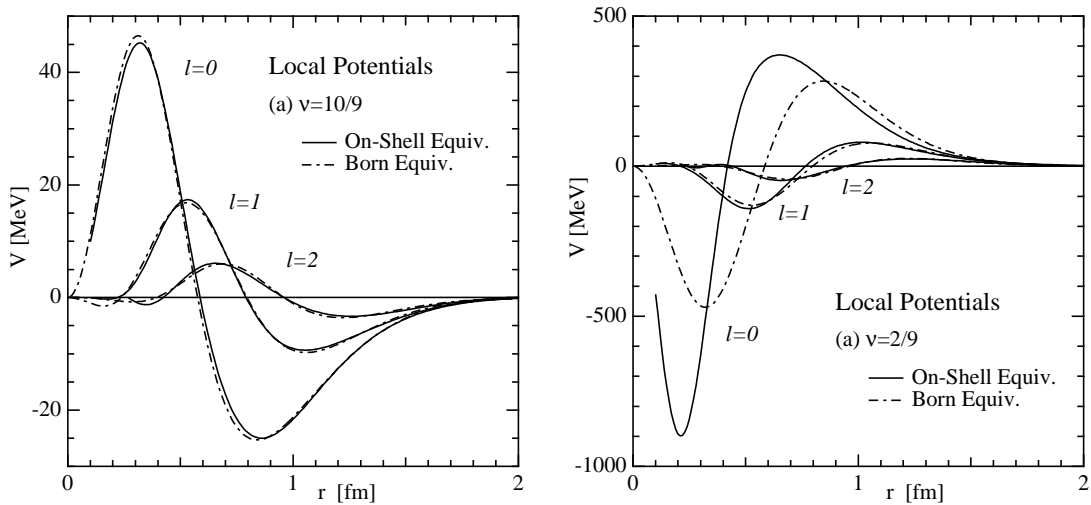}
	\caption{
On-shell-equivalent local potentials and Born-equivalent local potentials 
(see text).
(a) is for the $\nu=10/9$ case,
and (b) is for the $\nu=2/9$ case.
\label{fig2}
}
\end{figure*}
The scattering phase shift obtained from the nonlocal potential $V_{K}$
is shown in Figure \ref{fig1}
in solid lines.
For the $\nu=10/9$ case (Figure \ref{fig1}(a)), 
the phase shift is positive at 
the low energy region. It, however, goes negative 
above about $\beta k \sim \sqrt{\ell+{3\over 2}}$, the momentum of the $0\ell$ state.
When $\nu$ becomes small, there
appears an almost-forbidden resonance at this momentum.
In the $\nu=2/9$ system (Figure \ref{fig1}(b)), 
the phase shift is negative
at the low-energy region
while it becomes positive above a certain energy.
The reason is as follows.
The direct part of the kinetic energy mixes
the $0s$ and $1s$ states; when $\nu<1$, this mixing   
is suppressed by 
the potential in eq.\ (\ref{eq:KV0}).
Above the kinetic energy of the $0\ell$ state,
the potential becomes attractive,
which also comes from the suppression of the mixing.
On the other hand,
when $\nu > 1$, the mixing is enhanced.
Then, the potential becomes attractive in the low energy region
and repulsive in the high energy region.
The phase shift given by 
the lowest-order term, eq.\ (\ref{eq:KV0}), is also 
plotted in Figure \ref{fig1} in dotted lines. 
The full
 phase shifts are 
well-approximated by the 
lowest-order term.

The Born phase shifts of the
original nonlocal potential are plotted also in Figure \ref{fig1}
by dot-dashed lines.
The overall feature of the $\nu=10/9$ channel
is well reproduced by this approximation.
Since the potential is stronger, 
the Born approximation does not work well in the $\nu=2/9$ $\ell=0$ channel.
The momentum which gives ${\delta(k_{0})=0}$,  $k_{0}$, 
however, 
is well simulated.
As we mentioned before, this momentum 
is obtained as the solution of
$\phi_{n=1\ell}(k_{0})=0$ approximately. 
Here, $k_{0}\sim{3\over 2b}$ = 2.5 fm$^{-1}$
for the $\ell=0$ channel.

The wave functions obtained from this nonlocal 
potential as well as the ones from the Born-approximation
are plotted in Figure \ref{wfn1} for higher momentum
$k=2$, 2.5, and 3 fm$^{-1}$ 
($E=166$, 259, and 373 MeV, respectively.)
Those of the $\nu=10/9$ $\ell=0$ channel are shown in 
Figure \ref{wfn1}(a); the wave functions are very close to 
the free wave functions for all $k$.
On the other hand, 
the wave functions of the $\nu=2/9$ $\ell=0$ channel  
shown in Figure \ref{wfn1}(b) deviate largely from the free wave function.
At $k=2.5$ fm$^{-1}$, where the phase shift is almost zero,
there is a large enhancement of the wave function at the short range
region.
This comes from the attraction at 
short distance of the nonlocal potential.
This tendency can also be seen for the Born wave function.
The degree of the enhancement, however, is different
even where the Born phase shift is close to the original one;
the original wave function is much larger than that of the Born approximation 
at $k=2.5$ fm$^{-1}$, whereas it becomes smaller 
at $k=3$ fm$^{-1}$. The energy dependence of the original 
wave function is much larger than the Born wave function.

\subsection{Equivalent potentials}
\label{sec42}

In Figure \ref{fig2}, 
we plot the  on-shell-equivalent local potentials 
obtained for the 
$\nu=10/9$ (Figure \ref{fig2}(a)) and $\nu=2/9$ (Figure 
\ref{fig2}(b)) channels in solid lines.
The potential for $\nu=10/9$ is attractive at long range and
repulsive at short range.
Overall size, however, remains rather small.
The potential for the $\nu=2/9$ system has an
opposite sign and is very large.
Especially, the $\ell=0$ system
has an exceptional feature.
It has large potential barrier at the longer distance
with a deep attractive pocket in the short range.
There is a quasi-bound state in this 
pocket; the scattering wave function is orthogonal 
approximately to 
the state except for around the resonance energy.
To have such a pocket is the way to express an 
almost forbidden state 
by a local potential.

\begin{figure}
	\includegraphics*[scale=0.6]{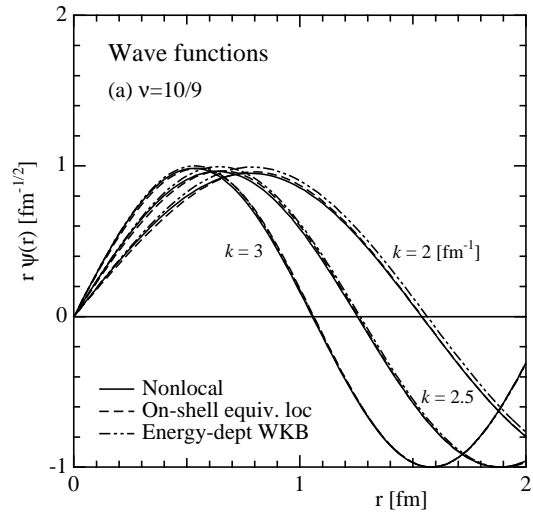}
	\caption{Wave functions obtained 
from the on-shell equivalent local potential
(dashed lines) and that from 
the original nonlocal potential (solid lines)
for the $\nu=10/9$ case.
The result from the energy-dependent potential
is also plotted by the double-dot-dashed line.
\label{wfn2a}
}
\end{figure}
\begin{figure}
	\includegraphics*[scale=0.6]{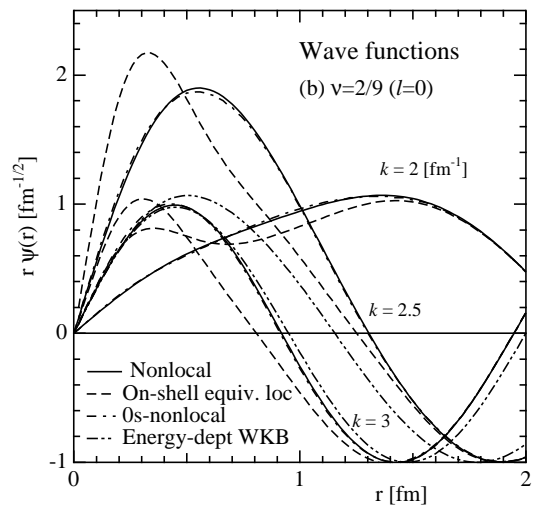}
	\caption{Wave functions obtained 
from the on-shell equivalent local potential
(dashed lines) and that from 
the original nonlocal potential (solid lines)
for the $\nu=2/9$ case.
That from the partially local potential 
defined by eq.\ (\ref{eq:0snloc}) (dot-dashed lines)
and that from the energy-dependent potential 
(double-dot-dashed lines) are also plotted. 
\label{wfn2b}
}
\end{figure}
\begin{figure}
	\includegraphics*[scale=0.6]{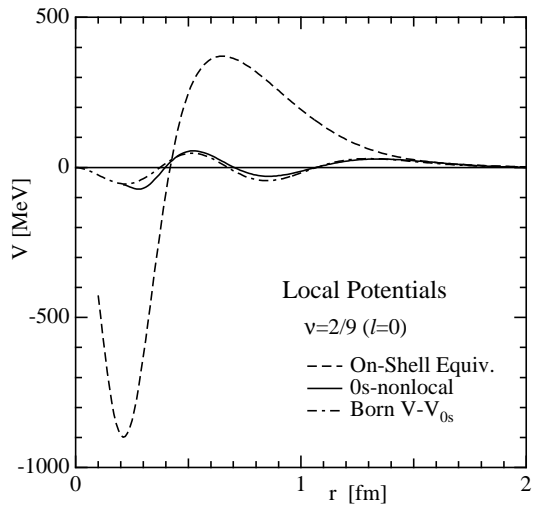}
	\caption{ Local part of the partially local potential,
$V^{0s{\rm -nonloc}}_{loc}(r)$,
for the $\nu=2/9$ case defined by eq.\ (\ref{eq:0snloc}).
The difference between the full and 0$s$-1$s$ part of 
the Born-equivalent local potential
is also plotted in the dot-dashed line.
\label{fig:V0snloc}}
\end{figure}
%

\begin{figure*}
	\includegraphics*[scale=0.6]{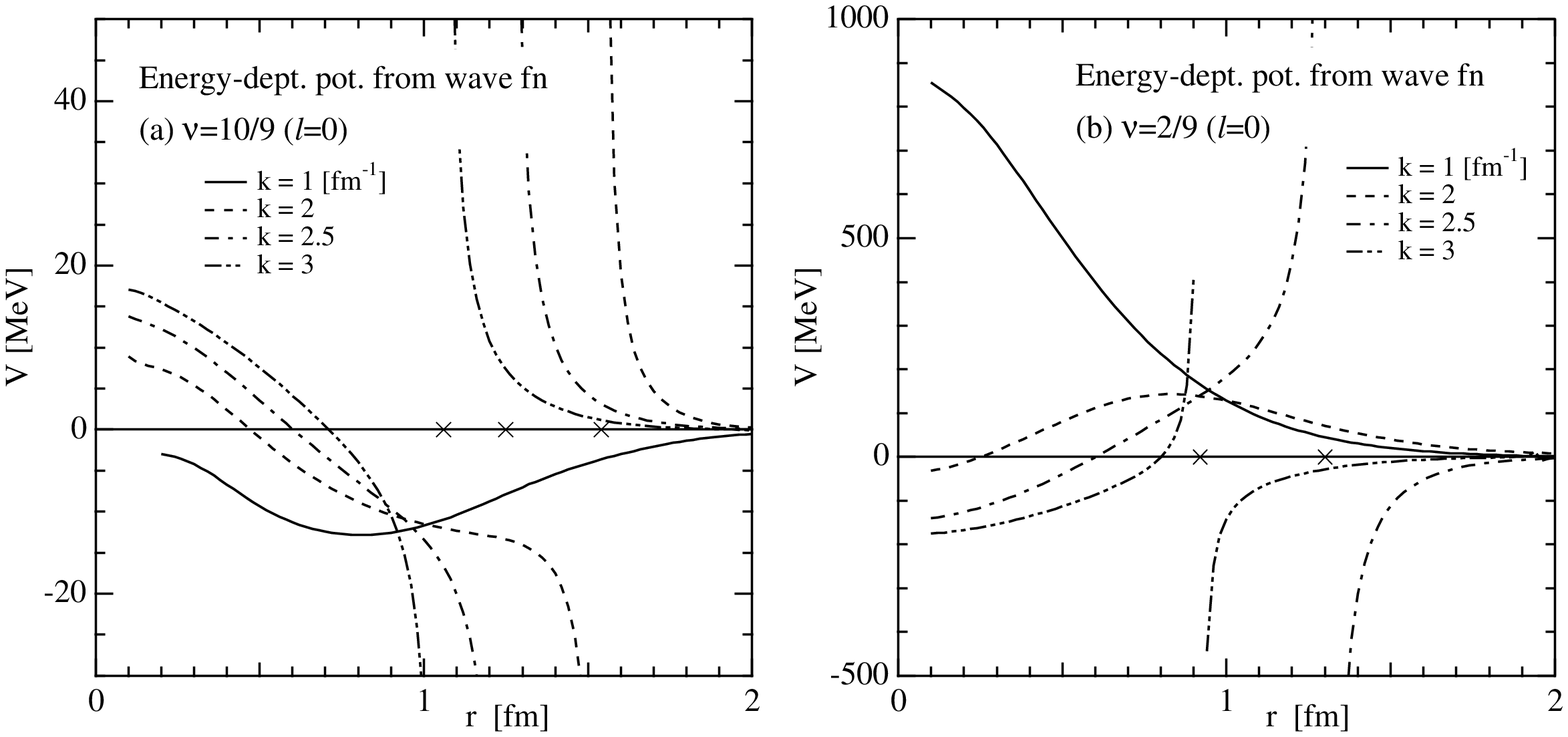}
	\caption{Energy-dependent
potentials by division (see text).
(a) is for the $\nu=10/9$ case,
and (b) is for the $\nu=2/9$ case.
Node points of the wave functions are marked by $\times$'s.
\label{vdiv}
}
\end{figure*}

The wave functions obtained from the on-shell-equivalent local 
potential
are plotted in Figures \ref{wfn2a} and  \ref{wfn2b}
for the $\ell=0$ wave.
It is found that the wave function
is similar to the original one for  the $\nu=10/9$ case.
 On the other hand,
the wave function of the system with an almost-forbidden state
has an artificial bump at the short distance at the low energy.

The Born-equivalent local potentials are plotted also in 
Figure \ref{fig2}.
They
are similar to the on-shell-equivalent local potentials
 except for the $\nu=2/9$ $\ell=0$ channel.
There, both of the barrier 
and the attractive pocket
of the on-shell-equivalent potential are more manifest.
As seen in Figure \ref{fig1}(b), Born approximation
is not valid in this channel.
Even so, the fact that those two potential deviate from each other
indicates the nonlocality of the original potential.
Though the size of the deviation is probably affected by the
Born approximation, 
it seems safe to conclude that the nonlocality is very large 
in this channel.

\subsection{Partially local potential}
\label{sec44}

We obtained the local part  in the
partially local potential,
$V^{0s{\rm -nonloc}}_{loc}(r)$,
defined by eq.\ (\ref{eq:0snloc}) 
for the $\nu=2/9$ $\ell=0$ channel.
The size of this local part becomes much smaller than 
the on-shell-equivalent local potential,
and is found to have more nodes
as seen in Figure \ref{fig:V0snloc}.
We also plot
the higher-order term of 
the Born-equivalent potential,
namely, the difference between 
the Born-equivalent potential of the full potential 
and that of the $n$ or $n'=0$ term.
This higher-order term is very similar to 
$V^{0s{\rm -nonloc}}_{loc}(r)$,
while the on-shell equivalent potential 
is very different from the Born-equivalent one.
This suggests that 
it is important to take 
the nonlocality
in the $n$ or $n'=0$ term into account.

The wave function corresponding to this 
partially local potential is also shown in  Figure \ref{wfn2b}.
The obtained wave function is similar to that of the nonlocal 
potential.
This confirms
that
the term between the $0s$ and the other states
is the dominant part of the nonlocal potential
and that the nonlocality can be taken care of 
by considering this term.
\begin{figure*}
\includegraphics*[scale=0.6]{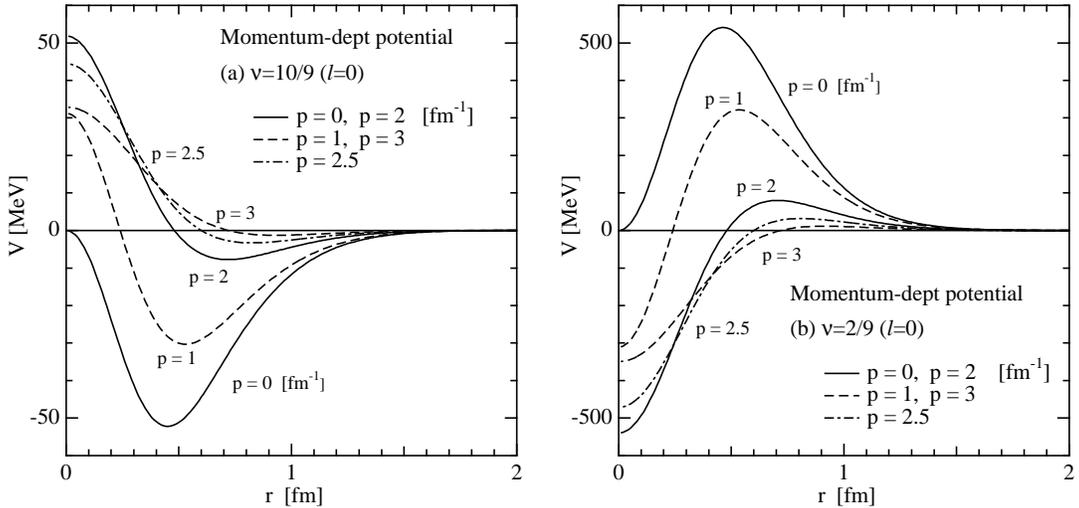}
\caption{Momentum-dependent
potentials, 
$U(r,p)$ (see text).
(a) is for the $\nu=10/9$ case,
and (b) is for the $\nu=2/9$ case.
\label{vpdepfig}}
\end{figure*}
\begin{figure*}
	\includegraphics*[scale=0.6]{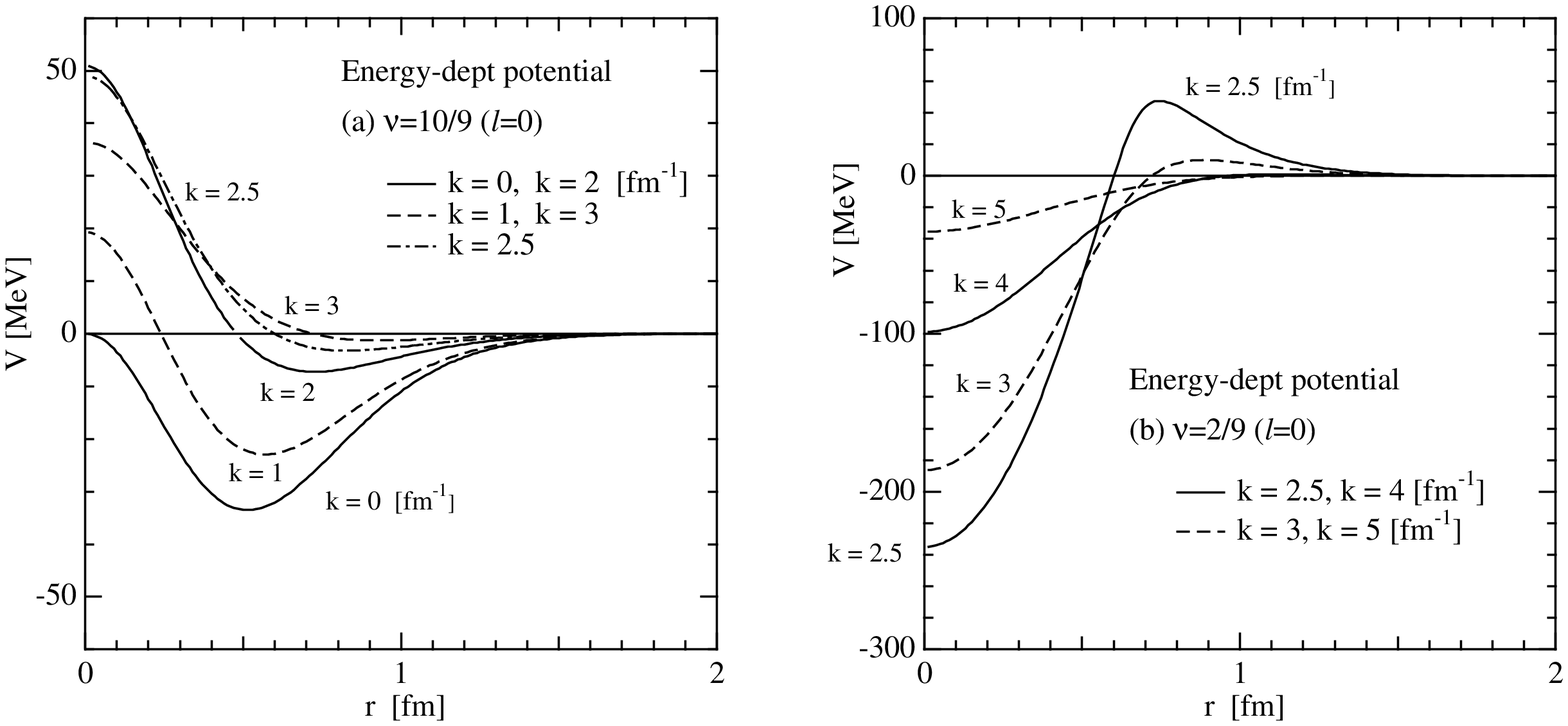}
	\caption{Energy-dependent
potentials with the WKB method,
 $U(r,E)$ (see text).
(a) is for the $\nu=10/9$ case.
(b) is for the $\nu=2/9$ case.
\label{vedep}}
\end{figure*}

\subsection{Energy-dependent potentials}
\label{sec45}

The energy-dependent potentials 
constructed from the $s$-wave wave functions
by eq.\ (\ref{eq:vdiv}) are plotted in Figure \ref{vdiv}.
As we discussed in the previous section, 
there is a singularity at each of the node points
of the wave function.
Also, numerical error is large at $r\sim 0$.
Nevertheless, one can see that,
for the $\nu=10/9$ channel,
the short range part increases as the energy increases.
As for the
$\nu=2/9$ channel,
a large repulsive core appears at the low energy region, 
which decreases rapidly as the energy increases.
Above $k\sim 2.5$ fm$^{-1}$, the short range repulsion
disappears.

The momentum-dependent potentials 
obtained by the WKB method, $U(r,p)$ in 
eq.\ (\ref{eq:p2E}),
are plotted in Figure \ref{vpdepfig}.
As seen from Figure \ref{vpdepfig}(a), $U(r,p)$ of the
 $\nu=10/9$ $\ell=0$ channel is negative definite at
$p=0$. As the momentum increases, 
a repulsion appears at the short range region.
$U(r,p)$ of the  $\nu=2/9$ $\ell=0$ channel 
is shown in Figure \ref{vpdepfig}(b), which is 
the same as the other channel except for the overall
amplitude and sign.

For the $\nu=10/9$ case,
by solving eq.\ (\ref{eq:p2E}) self-consistently,
we have the energy-dependent potentials,
$U(r,E)$, which are shown in Figure \ref{vedep}(a).
It shows similar behavior to the momentum-dependent potential.
At $E=0$ MeV, or $k=0$ fm$^{-1}$, 
it is simply attractive.
As the energy increases, however, the repulsive part appears
at the short range. 
As discussed in eq.\ (\ref{eq:U0rE}),
when $\beta^2 p^{2}$ is small, the potential at $r=0$
increases by ${16 (\sqrt{\nu}-1)\over 1+ 16 (\sqrt{\nu}-1)}E$,
which is 0.46$E$.
Though 
one has to add the contribution from the interaction
to see the energy dependence of the repulsive core of the 
two-baryon systems,
it is interesting to see that the short-range part of the 
potential may move by a large amount  as the energy increases.
Above about $k=3$ fm$^{-1}$, the potential
gradually decreases.
%

\begin{figure*}
	\includegraphics*[scale=0.6]{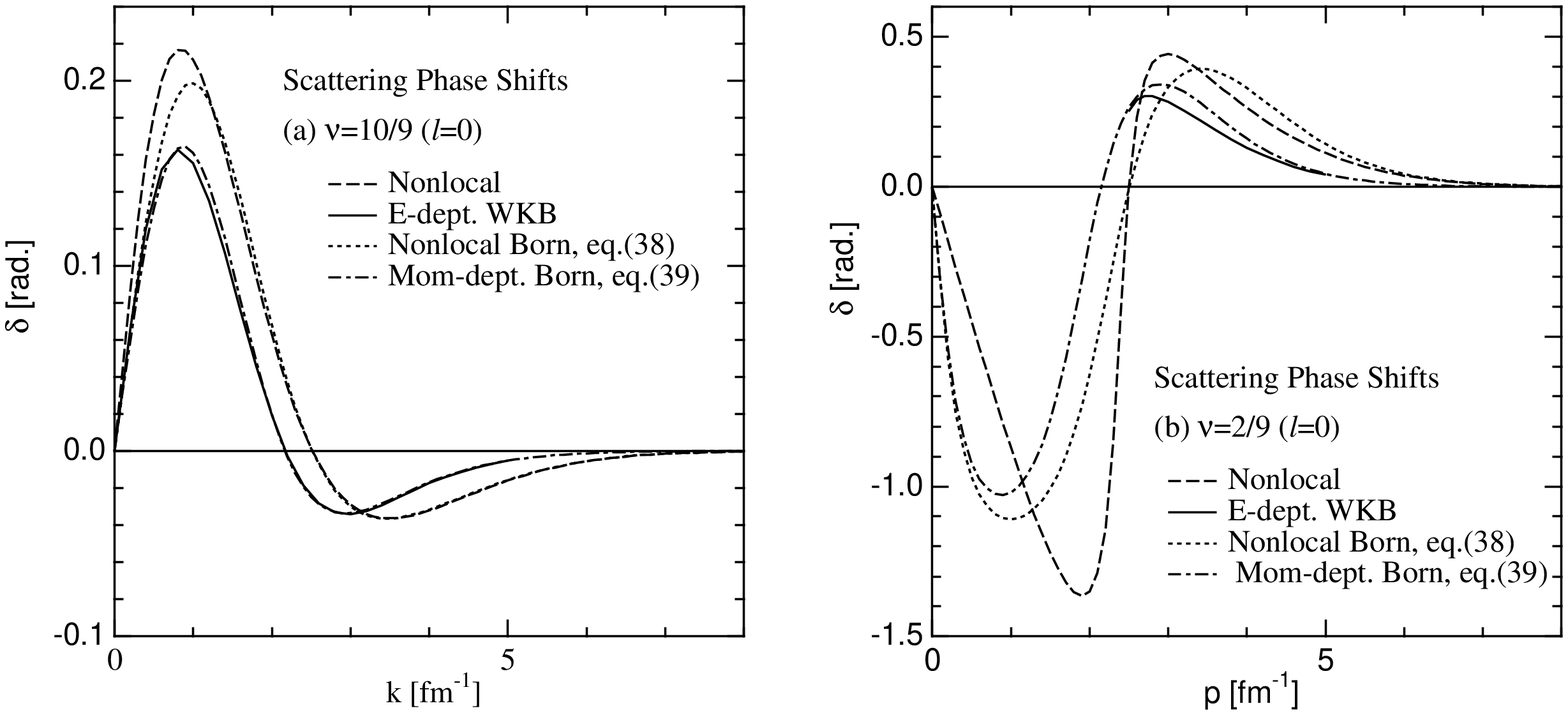}
	\caption{Scattering phase shifts obtained from the 
energy-dependent
potentials with the WKB method
as well as those of the momentum-dependent potentials with the
Born approximation.
(a) is for the $\nu=10/9$ case.
(b) is for the $\nu=2/9$ case, where that of the WKB method is
calculated only above $k\geq 2.5{\rm fm}^{-1}$.
\label{psvedep}}
	
\end{figure*}
The phase shift given by this energy-dependent local potential
from the WKB method 
is plotted in Figure \ref{psvedep}(a).
Overall feature is well simulated by this potential.
The momentum
where the sign of the phase shift changes, however,
becomes smaller.
The deviation from the original one 
comes from the approximation used to derive
eq.\ (\ref{eq:Edep2}) from
eq.\ (\ref{eq:EdepRL}), but not from the WKB method.
This is confirmed by employing the Born approximation
for each potential in eq.\ (\ref{eq:EdepRL}) and in
eq.\ (\ref{eq:Edep2}), which 
is shown also 
in Figure \ref{psvedep}.

We plot the wave function from the energy-dependent potential 
by WKB method for the $\nu=10/9$ case in Figure \ref{wfn2a}.
The wave function is similar to the original one
in spite of the deviation found in the phase shift, probably
because the size of the potential is rather small.

The $\nu=2/9$ potential shows more notable feature.
There are two real solutions to fill the conditions eq.\ (\ref{eq:p2E})
for each $r$ and $E$ at energy higher than $k=2.5$ fm$^{-1}$.
On the other hand, at energy lower than $k=2.5$ fm$^{-1}$, 
the solutions have a complex value around $r \sim 0.7$ fm.

In the Figure \ref{vedep}(b), we plot 
the solution which goes to zero 
at $r\rightarrow \infty$ for $k\geq 2.5$fm$^{-1}$.
At  $k=2.5$ fm$^{-1}$, the potential 
has attractive pocket in the short range region 
with a barrier around $r \sim 0.7$ fm.
Both of the barrier and the attractive pocket becomes smaller as
the energy increases.

The phase shifts for $k\geq 2.5$ fm$^{-1}$ of the $\nu=2/9$ case 
are shown for in Figure \ref{psvedep}(b).
The resonance also moves to a smaller momentum.
The deviation also seems to come from 
the approximation used to derive eq.\ (\ref{eq:Edep2}) from
eq.\ (\ref{eq:EdepRL}).

The wave function from the energy-dependent potential 
by WKB method for the $\nu=2/9$ case in Figure \ref{wfn2b}.
The wave function at $k=3$ fm$^{-1}$ is similar to the nonlocal one.
On the other hand, that of $k=2.5$ fm$^{-1}$ deviates largely
because the resonance occurs at the lower energy.

\section{Summary}
\label{sec5}

Pauli-Blocking effect on the kinetic term is investigated
by employing the 
quark cluster model. The effect can be expressed by 
a potential which is highly nonlocal. 
It is found that the Pauli-Blocking effect
does not cause a simple attraction or repulsion.
In order to see this point, we have calculated the phase shifts for two 
cases, namely, the one where the normalization factor $\nu$ 
(see eq.\ (\ref{eq:nu})) 
is larger than 1 and the other case where $\nu$  
is smaller than 1.  
In the first case ($\nu=10/9$, which corresponds 
to the two-nucleon $S$-wave channel), the phase shift is attractive at low energy 
region while it becomes repulsive as the energy increases. 
In the second case ($\nu=2/9$, which corresponds
to $\Sigma$N($I$=3/2,$S$=1)), the phase shift is repulsive at low energy 
region while it becomes attractive as the energy increases. 
These behaviors of the phase shifts can be understood in the following way. 
Since the Pauli-blocking changes the degrees of the mixing between the
incoming wave and the 0$\ell$ state of the inter-cluster
wave function, the behavior of the resulting effect 
above the energy of the 0$\ell$ state is different from that 
in the lower energy-region.

We also look into
the properties of this nonlocal potential by constructing 
four kinds of local potentials:
1) the one which gives the same scattering phase shift as that of the 
original potential,
2) the one which gives the same Born phase shift,
3) the one with energy dependent potentials obtained from the wave 
function,
and 4) the one with energy dependent potentials obtained by the WKB 
method.
It is found that the behavior of the equivalent local potentials 
depends strongly on the size of the 
Pauli-Blocking effect.

In the channel where the Pauli-blocking effect is 
small ($\nu=10/9$), 
the former two local potentials, which are very similar to each other,
simulate the nonlocal potential well even for the off-shell behavior. 
The energy dependent potentials given 
by the WKB method can also be obtained and found to increase
as the energy increases.

On the other hand, where the Pauli-blocking effect is large ($\nu=2/9$),
the off-shell behavior of the equivalent local potential is 
very different from the original one.
It is because the local potential simulates the almost-forbidden 
resonance by having a deep attractive pocket at the very short range
with a large barrier.
Therefore it is very difficult to simulate the on- and off-shell 
behavior of the  nonlocal potential simultaneously in terms of 
the local potential.  However, if we keep the main part 
of the nonlocal potential, namely, the term coming from the $0s$-$1s$ 
component, the rest terms can be nicely simulated by a local potential 
not only for the on-shell but also for the off-shell behaviors.  
We have also calculated the energy-dependent potential by the WKB method 
and found that it becomes
complex potential at some area.
For the higher energy, the obtained potential is real.
There we also find an attractive pocket and a barrier, both of which 
disappear
as the energy increases.

The preset work indicates that in a certain channel
the nonlocal treatment is essential.
The place where the off-shell features become important,
such as the baryons in nuclei,
it should be checked whether the use of a local potential
is appropriate or not.
For this purpose, nonlocality in the contribution from the 
interaction between quarks should also be considered.
It is also interesting to see
the situation in  coupled-channel systems,
which is now underway.

\bigskip

This work is supported in part by
the JSPS Grant-in-Aid for Scientific Research (C)(2)11640258
and (C)(2)12640290.

\section*{appendix}
Here we show how we expand
the  exchange term of the orbital normalization kernel,
$N_{\rm ex}(\tvecx,\tvecy)$.
This term has the form,
\begin{eqnarray}
N_{\rm ex}(\tvecx,\tvecy) 
&=& N_{0} \exp[-S(x^{2}+y^{2})+2T \tvecx\cdot\tvecy]~,
\label{eq:NexUV}
\end{eqnarray}
which can be expanded as
\begin{equation}
N_{0}
\sum_{n\ell m}\, \theta^{2n+\ell}
\psi_{n\ell m}(\beta,\tvecx)^{*}\psi_{n\ell m}(\beta,\tvecy)
\end{equation}
Here, $\psi_{n\ell m}(\beta,\tvecx)$ is the harmonic oscillator
wave function with the size parameter $\beta$.
$N_0$, $S$ and $T$ in eq.\ (\ref{eq:NexUV}) can be
expressed by using $\beta$ and $\theta$, as
\begin{eqnarray}
N_0 &=& \left\{ \pi \beta^{2}(1-\theta^{2}) \right\}^{-3/2}
 \\
S&=& {1+\theta^2\over 1-\theta^2} {1\over 2 \beta^2}
\\
T&=&{\theta\over 1-\theta^2} {1\over \beta^2}~.
\end{eqnarray}

Suppose we have two clusters which consists of $N_{1}$ and $N_{2}$
particles exchanging $n$ particles to each other.
When the configuration of the clusters can be expressed as $(0s)^{N_1}$ and 
$(0s)^{N_2}$, respectively, with the size parameter $b$,
we have
\begin{eqnarray}
\beta^2&=& {N_1+N_2 \over N_1 N_2 } b^2
\\
\theta&=&{ N_1 N_2 -n(N_1+N_2)\over N_1 N_2}~.
\end{eqnarray}

In the present study,
$N_1=N_2=3$ and $n=1$.
So, $\beta^2={2\over 3} b^2$ and $\theta=1/3$.

The exchange part of the kinetic kernel can be written as
\begin{eqnarray}
\lefteqn{
K_{\rm ex}(\tvecx,\tvecy) 
}&&\nonumber\\
&=& K_{0} \left[ -U(x^{2}+y^{2})+2V \tvecx\cdot\tvecy +W\right]
N_{\rm ex}(\tvecx,\tvecy)~,
\end{eqnarray}
where
\begin{eqnarray}
K_0 &=& {3\over 4 m b^2}
 \\
U&=& {1+3\theta^2\over (1-\theta^2)^2} {2\over 3\beta^2}
\\
V&=&{\theta(3+\theta^2)\over  (1-\theta^2)^2} {2\over 3\beta^2}
\\
W&=&{5-\theta^2\over 1-\theta^2}+N_1+N_2-4 ~.
\end{eqnarray}
Or, it can be expanded as
\begin{eqnarray}
\lefteqn{K_{\rm ex}(\tvecx,\tvecy) ~~=~~ (N_1+N_2-2)~K_{0}~N_{\rm ex}(\tvecx,\tvecy)}
&&\nonumber\\
&+&
\sum_{nn'\ell m} K^{nn'\ell}
\psi_{n\ell m}(\beta,\tvecx)\psi_{n'\ell m}^*(\beta,\tvecy) 
\end{eqnarray}
with
\begin{widetext}
\begin{eqnarray}
K^{nn'\ell} &=& {2K_{0}\over3}\left\{
\delta_{nn'}
\left(2 n + \ell + {3\over2} \right)
+(\delta_{nn'+1}+\delta_{nn'-1})
\sqrt{(n_{<}+1)\left(n_{<} + \ell + {3\over2} \right)}
 \right\}\theta^{2 n_{<} + \ell}
\end{eqnarray}
where $n_<$ corresponds to the smaller one between $n$ and $n'$. 
\end{widetext}


\newpage

\end{document}